\newcommand{\CLASSINPUTtoptextmargin}{1.9cm}
\newcommand{\CLASSINPUTbottomtextmargin}{4.3cm}

\ifdefined\ONECOLUMN
\documentclass[journal, 12pt, onecolumn, draftclsnofoot, letterpaper]{IEEEtran}
\else
\documentclass[conference, a4paper]{IEEEtran}
\fi

\IEEEoverridecommandlockouts

\usepackage{amsmath,amssymb,mathrsfs,amsbsy}
\usepackage{cite}
\usepackage{graphicx,color}
\usepackage[bookmarks=false]{hyperref}
\usepackage{psfrag}
\usepackage{subfigure}
\usepackage{url}
\usepackage{balance}
\usepackage{stfloats}

\usepackage{array}
\usepackage{fancyhdr}
\usepackage{float}
\usepackage{epsfig}
\usepackage[nomain,acronym,toc]{glossaries}
\usepackage{algorithm,algorithmic}
\usepackage{tikz,siunitx}
\usepackage{makecell}

\newtheorem{thm}{Theorem}

\renewcommand{\eqref}[1]{(\ref{#1})}
\definecolor{sblue}{RGB}{0,51,160}
\definecolor{seaBlue}{RGB}{0,105,148}

\ifCLASSINFOpdf
\else
\fi

\makeatother

\begin{document}
\title{
Exploiting Non-uniform Quantization for \\Enhanced ILC in Wideband Digital Pre-distortion
}
\author{Jinfei Wang$^{1}$, Yi Ma$^{1\dagger}$, Fei Tong$^2$, and Ziming He$^2$\\
{\small $^{1}$5GIC and 6GIC, Institute for Communication Systems, University of Surrey, Guildford, UK, GU2 7XH}\\
{\small $^{1}$Emails: (jinfei.wang, y.ma, r.tafazolli)@surrey.ac.uk}\\
{\small $^{2}$Samsung Cambridge Solution Centre, Cambridge, UK, CB4 0AE}\\
{\small $^{2}$Emails: (f.tong, ziming.he)@samsung.com}
}


\maketitle

\begin{abstract}	
In this paper, it is identified that lowering the reference level at the vector signal analyzer can significantly improve the performance of iterative learning control (ILC).
We present a mathematical explanation for this phenomenon, where the signals experience logarithmic transform prior to analogue-to-digital conversion, resulting in non-uniform quantization.
This process reduces the quantization noise of low-amplitude signals that constitute a substantial portion of orthogonal frequency division multiplexing (OFDM) signals, thereby improving ILC performance.
Measurement results show that compared to setting the reference level to the peak amplitude, lowering the reference level achieves $3$~dB improvement on error vector magnitude (EVM) and $15$~dB improvement on normalized mean square error (NMSE) for $320$~MHz WiFi OFDM signals.
\end{abstract}


\IEEEpeerreviewmaketitle

\section{Introduction}\label{secI}
Iterative learning control (ILC) has gained significant attention in digital pre-distortion (DPD) for identifying optimal input signals to the power amplifier (PA) for its excellent robustness to the memory effects and non-linearity of PAs \cite{9337994,9575024,9322378,9360062,10018540}. 
However, the performance of ILC is highly dependent on the signal-to-noise ratio (SNR) of the captured PA output signals \cite{7885077}.
This challenge is particularly critical for wideband signals with pronounced memory effects.

In this paper, it is identified that lowering the reference level of the vector signal analyzer (VSA) can significantly improve ILC performance.
We present a mathematical explanation for this phenomenon.
When the reference level is lower than the peak signal amplitude, the signals experience logarithmic transform before passing the analogue-to-digital converter (ADC), resulting in non-uniform quantization.
This reduces the quantization step size for low-amplitude signals while relaxing it for high-amplitude signals, thereby reducing the quantization noise for low-amplitude components.
This property is particularly beneficial for orthogonal frequency-division multiplexing (OFDM) signals, as small-amplitude components constitute a substantial portion of OFDM signals.

Measurement results show that lowering the reference level by more than $17$~dB compared to the peak amplitude yields a $3$~dB improvement on error vector magnitude (EVM) and a $15$~dB improvement on normalized mean square error (NMSE) for WiFi OFDM signals with bandwidths of $160$~MHz and $320$~MHz.
Specifically, the ILC achieves $-49\sim-46$~dB EVM and up to $-54$~dB NMSE, demonstrating competitive performance compared to the literature (e.g., \cite{10018540,9337994,9575024,9322378,9360062}). 
Moreover, employing generalized memory polynomial (GMP) models to fit the ILC results achieves comparable EVM performance to the ILC and $-46\sim-43$~dB NMSE.
These findings confirm that lowering the VSA reference level is a simple yet effective approach to enhancing ILC performance, particularly for wideband OFDM signals.

\section{Principle of ILC}
Consider an OFDM signal $\mathbf{s}\in\mathbb{C}^{N\times1}$.
Ideally, we hope the output of the PA output $\mathbf{y}\in\mathbb{C}^{N\times1}$ is a linearly amplified version of $\mathbf{s}$.
However, due to the inherent non-linearity and memory effects of the PA, this ideal behavior is not easily achieved in practice. 
The aim of DPD design is to mitigate these imperfections.
Denote the PA function by $f(\cdot)$ and the DPD function by $g(\cdot)$.
Then, the target of DPD is to find $g(\cdot)$ such that
\begin{equation}
	\mathbf{y}=f(g(\mathbf{s}))=G\mathbf{s},
\end{equation}
where $G>1$ stands for the PA gain.

Since the PA function $f(\cdot)$ is generally unknown, it is challenging to directly determine $g(\cdot)$.
Various methods have been proposed to handle this issue, e.g., \cite{Eun1997,9825005,He2022}.
Among these, ILC has emerged as a promising approach, achieving impressive performance without requiring explicit modeling of $f(\cdot)$ or $g(\cdot)$ \cite{7522645}.
The idea of ILC is to directly acquire PA input signal $\mathbf{x}$ through iterative feedback such that 
\begin{equation}
	f(\mathbf{x})=G\mathbf{s}.
\end{equation}

To facilitate the discussion in the subsequent section, the principle of ILC with linear update is elaborated as follows.
Note that the discussion in the next section is also applicable to the gain-inverse update.
Denote the PA input and output of the $i$-th ILC iteration by $\mathbf{x}_i$ and $\mathbf{y}_i$ ($i=0,1,\cdots$), respectively.
The process begins with the initial PA input:
\begin{equation}
	\mathbf{x}_0=\mathbf{s}.
\end{equation}
In each iteration, the ILC signal is updated as
\begin{equation}
	\mathbf{x}_{i+1}=\mathbf{x}_i+\mathbf{e}_i,
\end{equation}
where the update error $\mathbf{e}_i$ is defined as
\begin{equation}\label{eq05}
	\mathbf{e}_i\triangleq\mathbf{y}_i-\mathbf{s}.
\end{equation}

According to \cite{7522645}, we will have the following equation after enough iterations:
\begin{equation}
	\lim\limits_{n\to\infty}f(\mathbf{x}_i)=G\mathbf{s}.
\end{equation}

In practice, $\mathbf{y}_i$ usually converge close to $G\mathbf{s}$ within $10$ iterations \cite{10018540,7885077}.

\section{Non-Uniform Quantization for ILC Feedback}\label{secIII}

The convergence analysis of ILC in \cite{7522645} was based on the assumption that $\mathbf{y}_i$ is exactly the PA output.
In practice, $\mathbf{y}_i$ is captured by the VSA before the update \eqref{eq05} happens.
This process involves multiple parameters, such as ADC quantization, ADC sampling rate, fractional delay compensation and thermal noise. 
In this paper, we focus on the mathematical explanation using the ADC quantization for the ILC performance with respect to the reference level.

Our testbed measurements in Section~\ref{secIV} reveal that setting the reference level below the peak signal amplitude can significantly improve ILC performance, even though it is commonly recommended to set the reference level above the peak signal amplitude.
It is likely that the signals undergoes a logarithmic transform prior to uniform quantization \cite{Pavan2016}.
Motivated by these observations, we aim to provide a mathematical explanation for the measurement results.

Assume that the quantization process for the in-phase and quadrature part of of $\mathbf{y}_i$ is performed independently, and the quantization of positive and negative values is symmetric \cite{Pavan2016}.
Hence, we are interested in the quantization of $|\Re(\mathbf{y}_i)|$, where $\Re(\cdot)$ stands for the real part of a scalar/vector, $|\cdot|$ for the absolute value of a scalar/vector.
This quantization problem can be linearly scaled to the quantization of $\mathbf{z}_i$:
\begin{equation}
	z_i(n)=|\Re(y_i(n))|/y_{i,\max},
\end{equation}
where
\begin{equation}
	y_{i,\max}=\max_n|\Re(y_i(n))|.
\end{equation}

It is trivial that $z_i(n)\in[0,1]$.
Consider this interval to be divided into $(Q+1)$ intervals ($Q\in\mathbb{N}^+$). 
Denote the $q$-th quantization thresholds by $d_q$ ($q=1,\cdots Q$).
The quantization step, denoted by $\delta_q$, is defined as
\begin{equation}
	\delta_q\triangleq d_{q+1}-d_{q},~q=1,2,\cdots Q-1.
\end{equation}
In the case of uniform quantization, we have
\begin{equation}\label{eq6789}
	d_q = q/(Q+1),~\delta_q=1/(Q+1).
\end{equation}

Mathematically, we are looking for a logarithmic transform function $h(z_i(n))$, where uniform quantization is applied to $h(z_i(n))$.
To facilitate our analysis, denote the ratio of $y_{i,\max}$ over current reference level by $\rho$ ($\rho>1$).
$h(z_i(n))$ should satisfy the following two conditions:
1) $\delta_q<\delta_{q+1}$; 2) $\delta_q$ decreases with the increase of $\rho$ for low-amplitude signals.
This leads to the following theoretical result.

\begin{thm}\label{thm01}
	A logarithmic transform that satisfy the aforementioned two conditions is 
	\begin{equation}\label{eq08}
		h(z_i(n)) \triangleq \log_{\rho}\left(\left(\rho-1\right)z_i(n)+1\right).
	\end{equation}
\end{thm} 
\begin{IEEEproof}
	$h(z_i(n))$ in \eqref{eq08} is monotonically increasing with $z_i(n)$.	
	Since $z_i(n)\in[0,1]$, it follows that $h(z_i(n))$ falls in $[0,1]$ as well.
	Then, when uniform quantization is applied to $h(z_i(n))$, the quantization thresholds are already given by \eqref{eq6789}.
	
	To know the equivalent quantization threshold $d_q$ for $z_i(n)$, we have the following equation:
	\begin{equation}\label{eq09}
		q/(Q+1)=\log_{\rho}\left(\left(\rho-1\right)d_q+1\right).
	\end{equation}
	Solving \eqref{eq09} yields
	\begin{equation}
		d_q=(\rho^{q/(Q+1)}-1)/(\rho-1).
	\end{equation}
	Hence, we have
	\begin{IEEEeqnarray}{rl}
		\delta_q&=\frac{\rho^{q/(Q+1)}(\rho^{1/(Q+1)-1})}{\rho-1}\\
		&=\frac{\rho^{q/(Q+1)}}{\sum_{k=1}^{k=Q}\rho^{k/(Q+1)}}.\label{eq12}
	\end{IEEEeqnarray}
	
	Since the numerator in \eqref{eq12} increases with the increase of $q$, we have $\delta_q<\delta_{q+1}$.
	{Moreover, as $\rho$ increases, $\rho^{k/(Q+1)}$ is increasing faster with the increase of $k$.
		Therefore, $\delta_q$ decreases with the increase of $\rho$ when $q$ is small, i.e., for low amplitude signals.}
	\textit{Theorem~\ref{thm01}} is proved.
\end{IEEEproof}

\begin{table}[t!]
	\center
	\caption{Testbed Parameters}
	\label{tabI}
	\begin{tabular}{c||c}
		\hline
		Parameter	& Value \\
		\hline
		DAC rate & $960$~MHz \\
		\hline
		ADC rate & $640$~MHz \\
		\hline
		DAC resolution & $12$-bit \\
		\hline
		ADC resolution & $12$-bit \\
		\hline
		Carrier Frequency & $5.32$~GHz  \\
		\hline
		OFDM Signal Bandwidth & $160$~\&~$320$~MHz  \\
		\hline
	\end{tabular}
\end{table}

\begin{figure}
	\centering
	\includegraphics[scale=1.5]{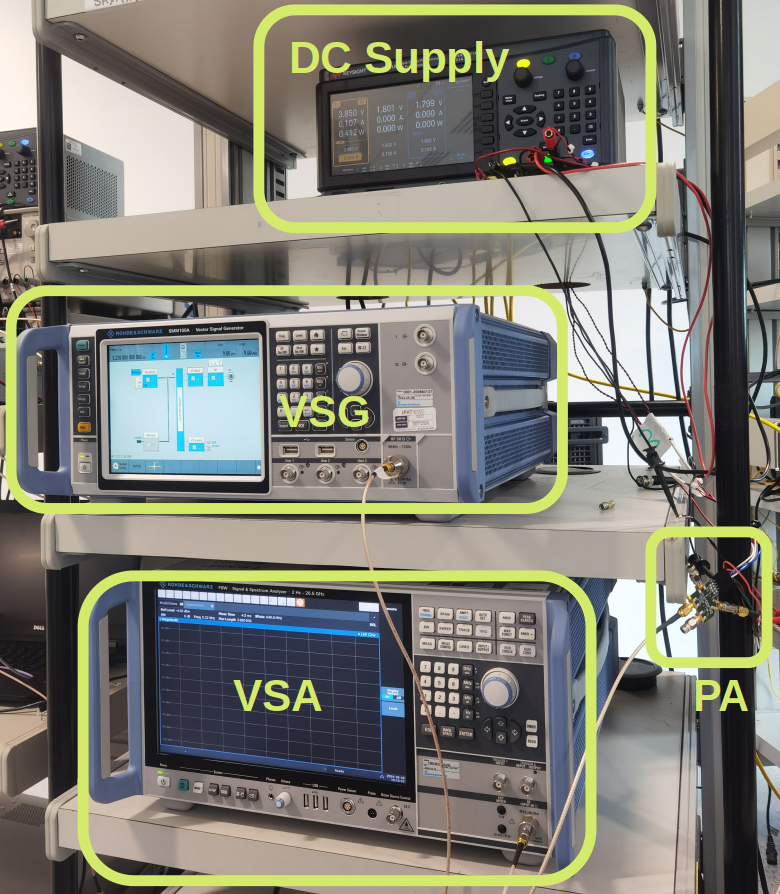}
	\caption{Testbed setup.}
	\vspace{-1.0em}
	\label{fig1}
\end{figure}

\textit{Theorem~\ref{thm01}} provides a theoretical explanation for phenomenon to be demonstrated later in Section~\ref{secIV}.
By introducing smaller quantization step sizes for low-amplitude signals, the associated quantization noise is reduced \cite{Liu2021}. 
This results in an improved SNR for the ILC feedback of OFDM signals, thereby enhancing ILC performance.

Although \textit{Theorem~\ref{thm01}} could not be explicitly verified in the VSA, it also provides theoretical insight for the design of VSA to improve the ILC performance, and consequently improves DPD performance.

\section{Measurement Results and Discussion}

\begin{figure}
	\centering
	\includegraphics[scale=0.46]{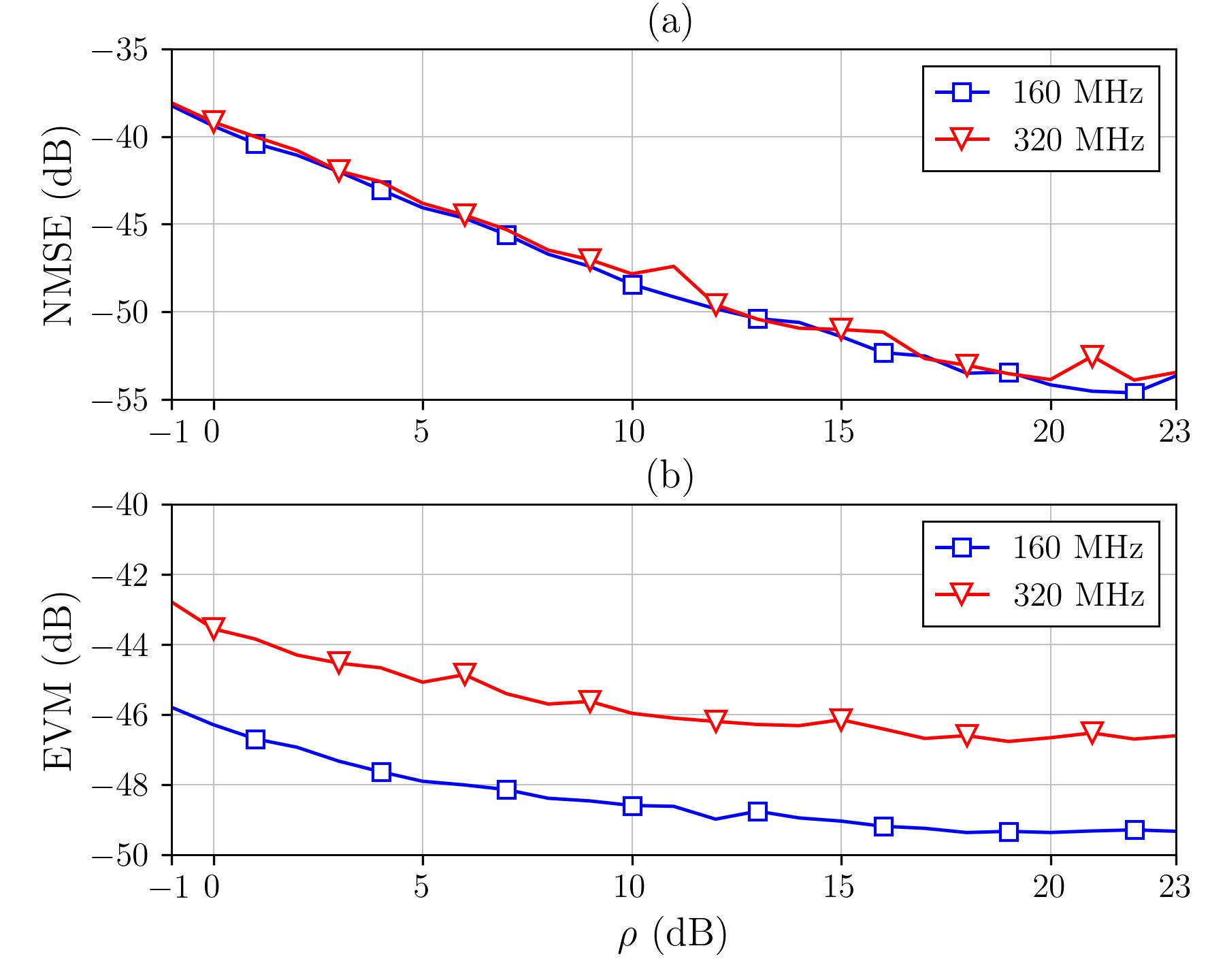}
	\caption{NMSE and EVM of ILC with respect to the ratio of signal peak amplitude over reference level (i.e., $\rho$).}
	\vspace{-1.0em}
	\label{fig2}
\end{figure}

\begin{figure}
	\centering
	\includegraphics[scale=0.46]{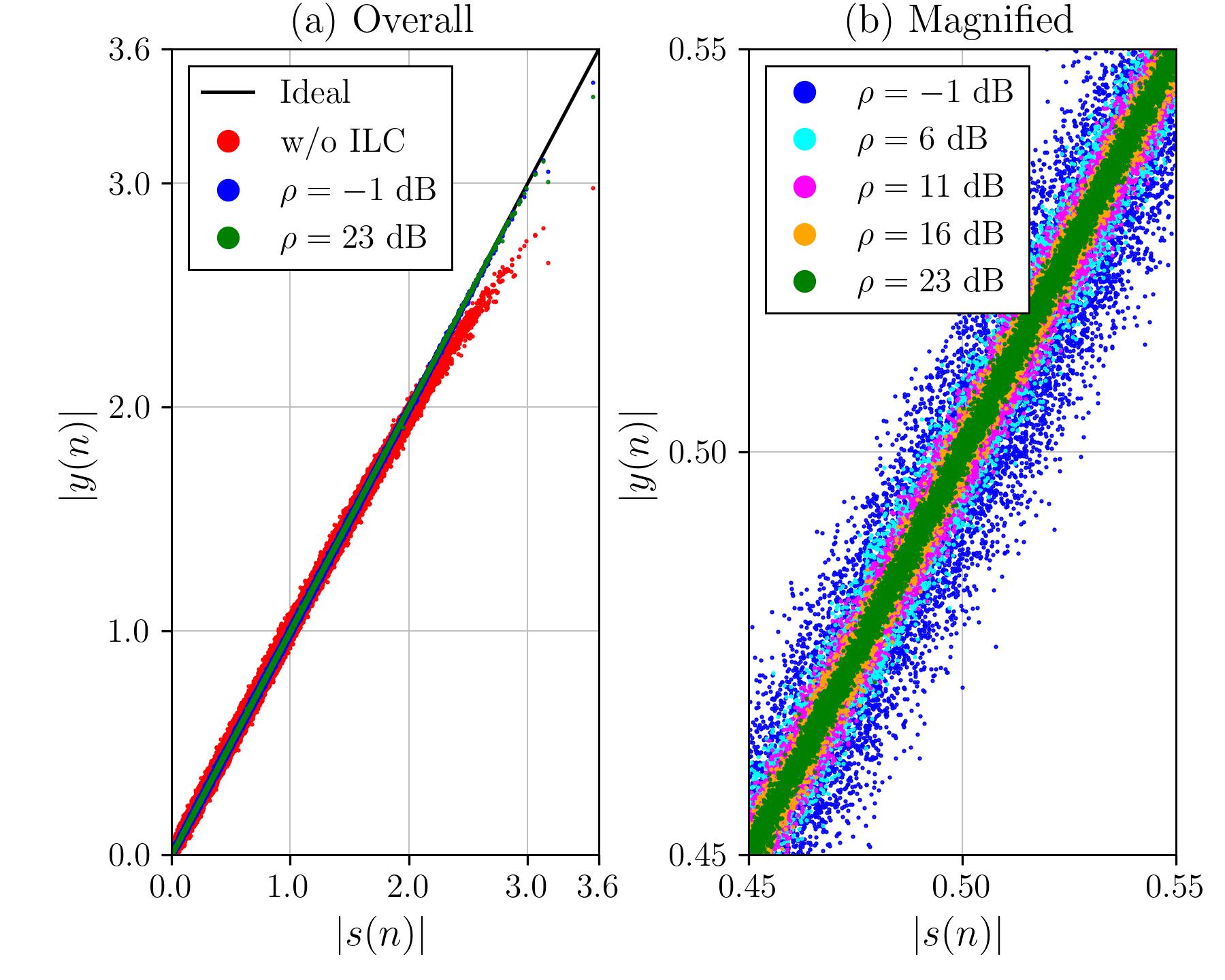}
	\caption{Overall and magnified AM-AM behavior of ILC with respect to the change of $\rho$ for $320$~MHz WiFi OFDM signals.}
	\vspace{-1.0em}
	\label{fig3}
\end{figure}

In this section, we present the testbed setup and the results of our experiments. The testbed consists of a Rhode \& Schwarz SMM100A vector signal generator (VSG), a Rhode \& Schwarz FSW VSA, and a Qorvo QM45500 PA in low-power mode, as shown in Fig.\ref{fig1}.
Detailed testbed parameters are provided in Table\ref{tabI}. 
The WiFi OFDM signals used in the experiments have a peak-to-average power ratio (PAPR) of $11$~dB.

\begin{figure}
	\centering
	\includegraphics[scale=0.46]{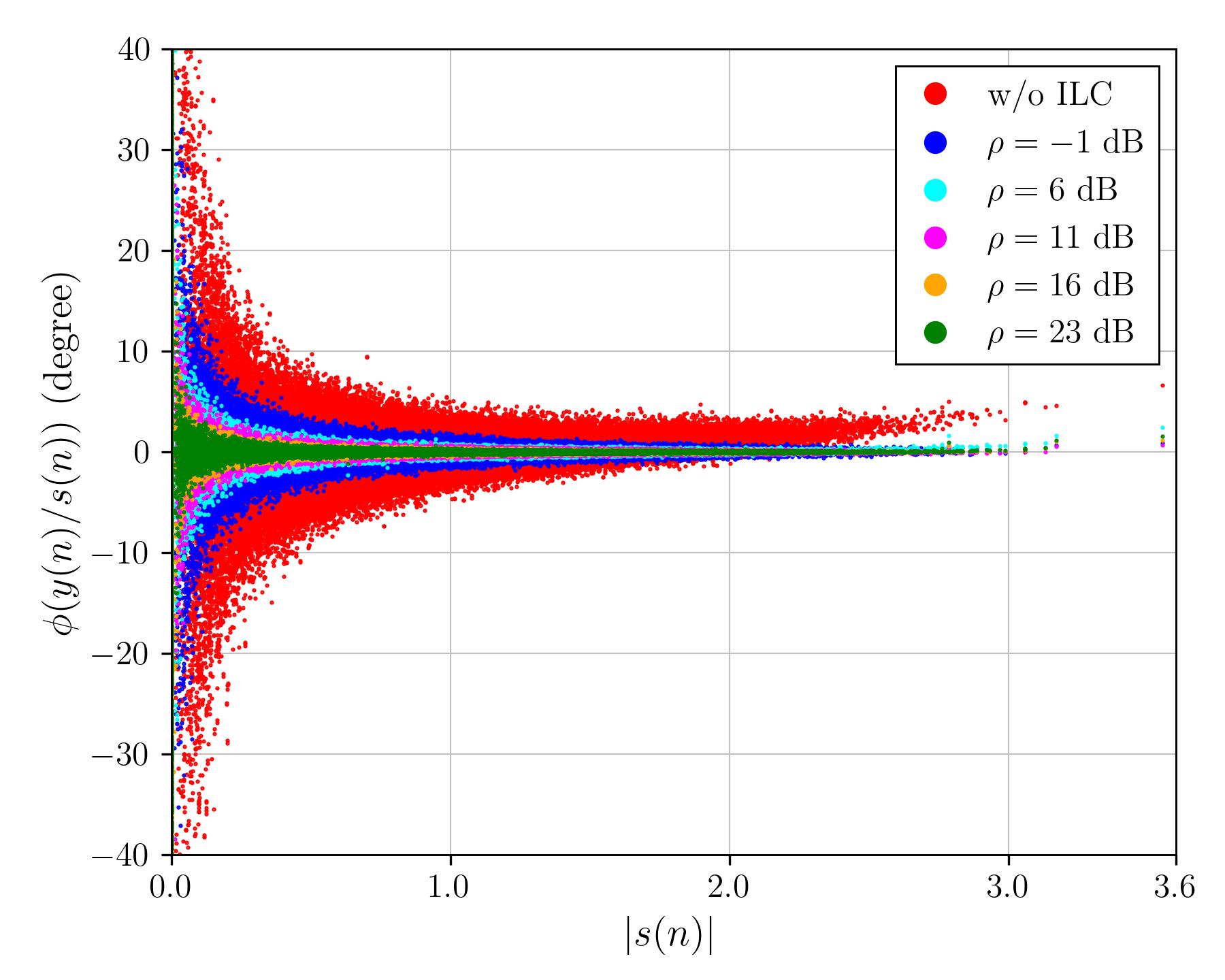}
	\caption{AM-PM behavior of ILC with respect to the change of $\rho$ for $320$~MHz WiFi OFDM signals.}
	\vspace{-1.0em}
	\label{fig4}
\end{figure}

\textit{Experiment~1:} The aim of this experiment is to demonstrate the change of ILC performance with respect to the VSA reference level, as shown in Fig.~\ref{fig2}$\sim$Fig.~\ref{fig4}.

Fig.~\ref{fig2} demonstrates the NMSE and EVM of ILC as the ratio of peak signal amplitude over the reference level $\rho$ increases from $-1$~dB to $23$~dB.
We start the test from $\rho=-1$~dB as this is the commonly suggested setup: the reference level is slightly higher than the peak signal amplitude.
Both the NMSE and EVM, increasing $\rho$ significantly improves the NMSE for $15$~dB and EVM for $3$~dB as shown in Fig.~\ref{fig2}(a) and Fig.~\ref{fig2}(b), respectively.

The NMSE performance in Fig.~\ref{fig2}(a) shows some swinging behavior. 
This is likely due to the resolution of fractional delay compensation is $32$ in our experiment.
This may not be fully sufficient for NMSE of $-50$~dB or less.

Fig.~\ref{fig3} shows the AM-AM behavior with respect to the change of $\rho$.
Specifically, Fig.~\ref{fig3}(a) shows the overall AM-AM behavior, where changing $\rho$ does not seem to introduce significant difference.
But when the AM-AM behavior is magnified, as in Fig.~\ref{fig3}(b), it is observed that increasing $\rho$ significantly reduces the memory effects for low-amplitude signals.
Similar phenomenon is observed in the AM-PM behavior, as shown in Fig.~\ref{fig4}.

It is also observed that when $\rho=16$~dB, it already achieves close performance to the case when $\rho=23$~dB in Fig.~\ref{fig2}$\sim$Fig.~\ref{fig4}.
In \textit{Experiment~2}, we will use the setting that $\rho=16$~dB.

\begin{figure}
	\centering
	\includegraphics[scale=0.46]{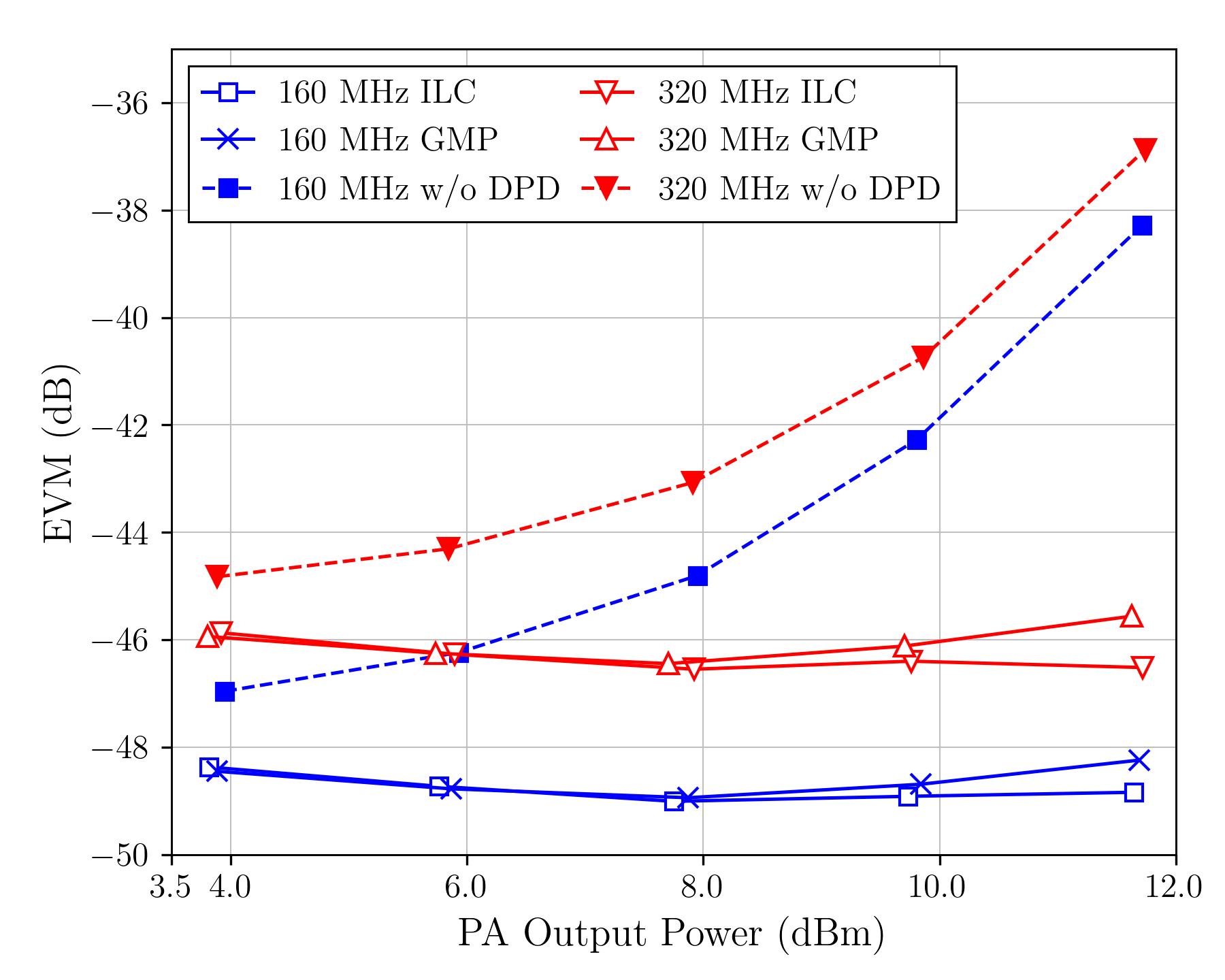}
	\caption{EVM of ILC and GMP with the increase of PA output power for $160$~MHz and $320$~MHz WiFi OFDM signals.}
	\vspace{-1.0em}
	\label{fig5}
\end{figure}

\textit{Experiment~2:}
The aim of this experiment is to demonstrate the DPD performance when using GMP (see \cite{Morgan2006}) to fit ILC results.
To this end, the least-square fitting is adopted \cite{Morgan2006}.
The experiment is based on $100$~waveforms in the training dataset as well as $100$~waveforms in the test dataset.

For the GMP, the maximum memory depth is $20$; the cross-memory depth is $1$. 
The polynomial order gradually increases from $3$ to $9$ with the increase of PA output power.

\begin{figure}
	\centering
	\includegraphics[scale=0.46]{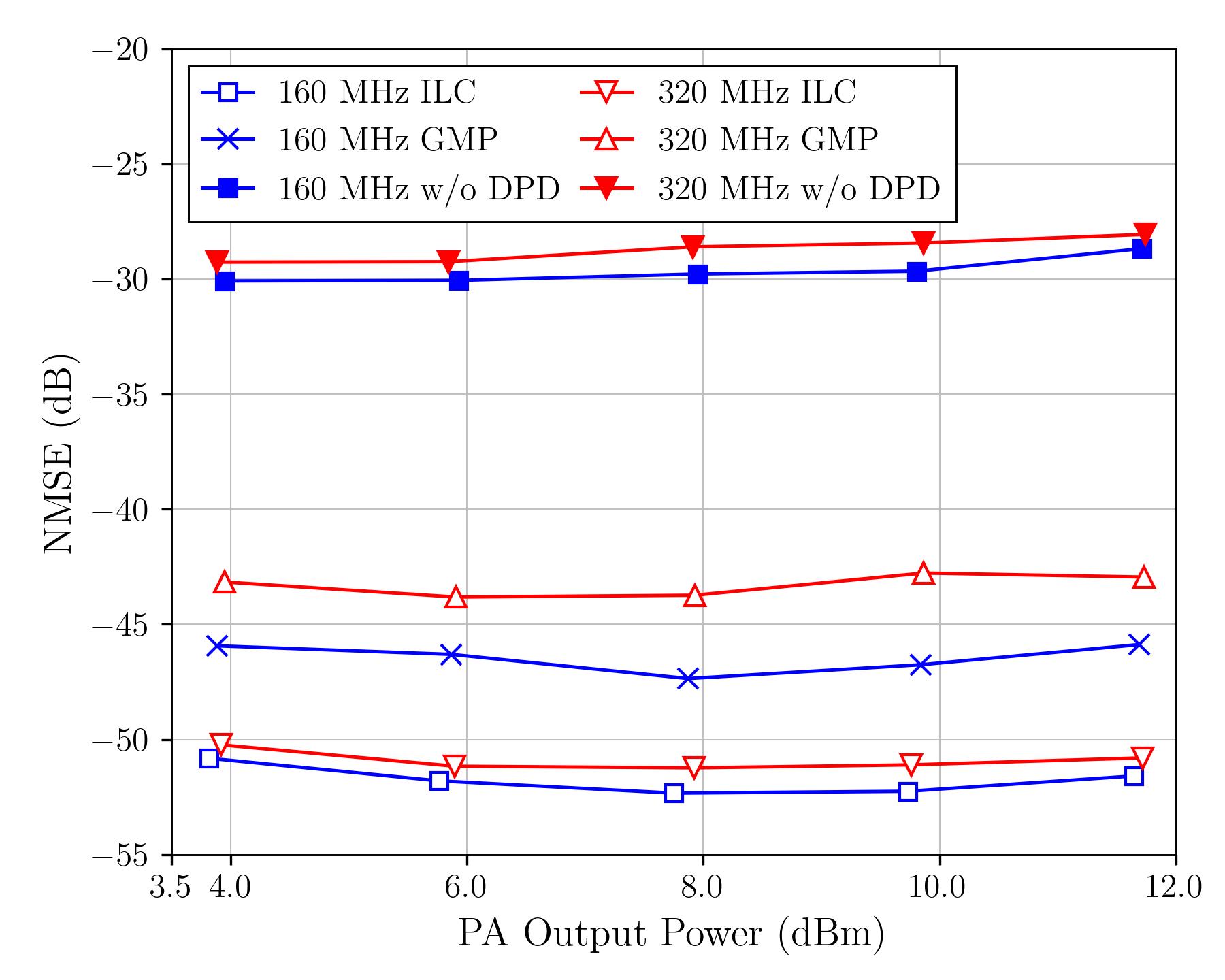}
	\caption{NMSE of ILC and GMP with the increase of PA output power for $160$~MHz and $320$~MHz WiFi OFDM signals.}
	\vspace{-1.0em}
	\label{fig6}
\end{figure}

Fig.~\ref{fig5} demonstrates the EVM as the PA output power increases. 
The GMP achieves almost identical EVM performance to the ILC (less than $1$~dB performance degradation).
For the NMSE, as shown in Fig.~\ref{fig6}, there is around $5$~dB degradation compared to the ILC.
But the GMP still achieves $-43$~dB NMSE.
This shows the performance improvement of ILC effectively transforms into the improvement on DPD performance.
The average NMSE in Fig.~\ref{fig6} is worse than Fig.~\ref{fig2}, this is due to the swinging phenomenon noted in Fig.~\ref{fig2}, which may stem from insufficient fractional delay compensation.

\begin{figure}
	\centering
	\includegraphics[scale=0.46]{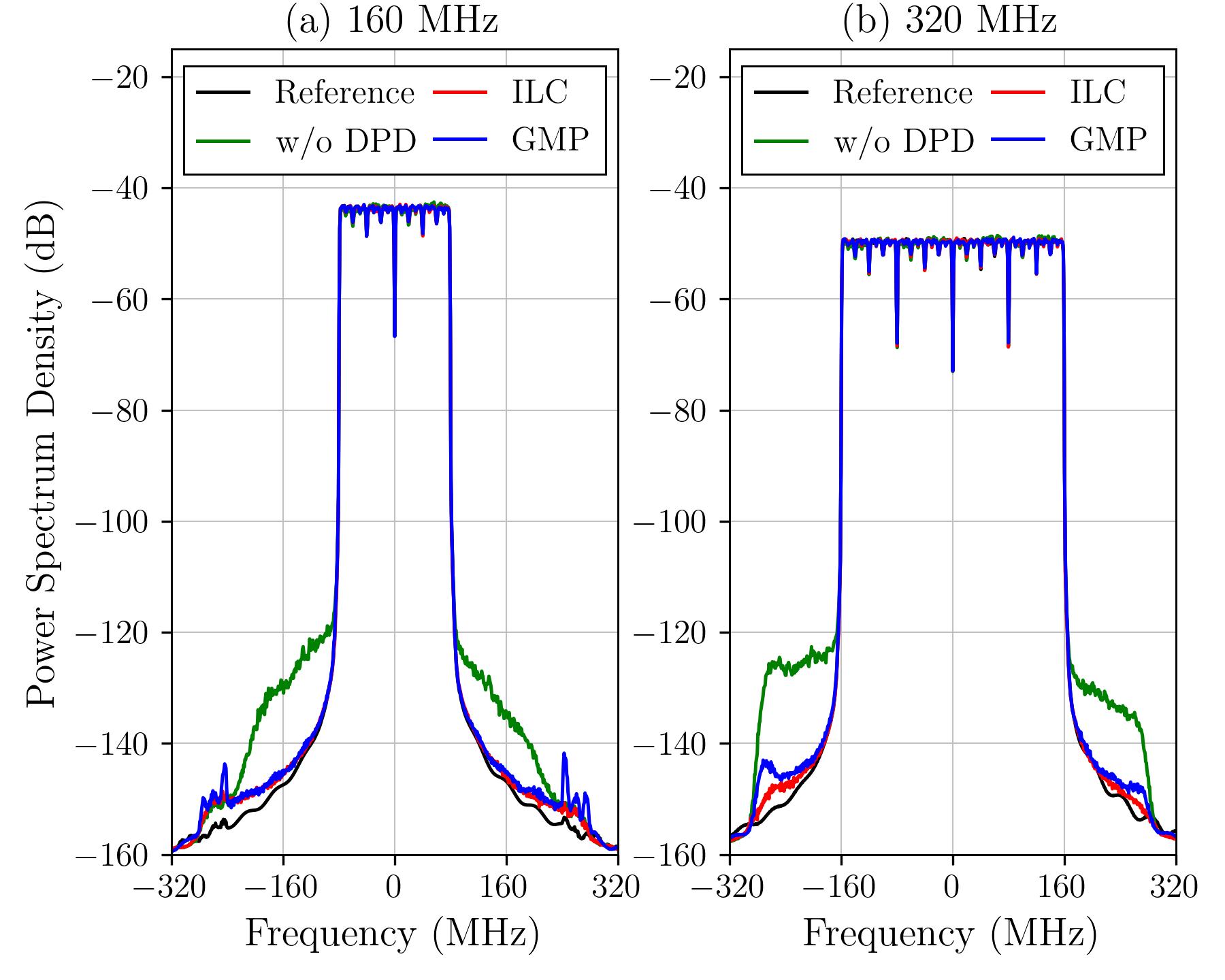}
	\caption{PSD of ILC and GMP when the PA output power is $11.8$~dBm for $160$~MHz and $320$~MHz WiFi OFDM signals.}
	\vspace{-1.0em}
	\label{fig7}
\end{figure}

Finally, Fig.~\ref{fig7} shows the power spectrum density (PSD) of ILC and GMP.
Both the GMP and ILC are close to the reference signal, with limited out-of-band emissions.
Notably, ILC still achieves good performance when the ADC rate is limited, as in Fig.~\ref{fig7}(b).
This is encouraging for the application of ILC in wideband DPD.

\section{Conclusion}\label{secIV}
In this paper, it has been identified that lowering the reference level at the VSA significantly enhances ILC performance.
A mathematical explanation has been provided for this phenomenon, where the signals undergo a logarithmic transformation prior to uniform quantization.
This effectively reduces the quantization steps as well as quantization noise for low-amplitude signals.
Measurement results have shown that lowering the reference level improves the NMSE by $15$~dB and the EVM by $3$~dB for $160\sim320$~MHz WiFi OFDM signals, achieving up to $-55$~dB NMSE and $-49\sim-46$~dB EVM.
These improvements effectively transforms into DPD performance enhancement when employing the GMP fitting model.

\section*{Acknowledgement}
This work is funded by Samsung-Academy Global Collaborative Research Programme.
Our sincere appreciation to Zern C. Tay and Jacob Sharpe for their outstanding support on testbed measurements.



\ifCLASSOPTIONcaptionsoff
\newpage
\fi

\bibliographystyle{IEEEtran}
\bibliography{bib/digitalPreDistortion,bib/Books_and_Standards,bib/GroupPaper}

\begin{thebibliography}{10}
\providecommand{\url}[1]{#1}
\csname url@samestyle\endcsname
\providecommand{\newblock}{\relax}
\providecommand{\bibinfo}[2]{#2}
\providecommand{\BIBentrySTDinterwordspacing}{\spaceskip=0pt\relax}
\providecommand{\BIBentryALTinterwordstretchfactor}{4}
\providecommand{\BIBentryALTinterwordspacing}{\spaceskip=\fontdimen2\font plus
\BIBentryALTinterwordstretchfactor\fontdimen3\font minus
  \fontdimen4\font\relax}
\providecommand{\BIBforeignlanguage}[2]{{%
\expandafter\ifx\csname l@#1\endcsname\relax
\typeout{** WARNING: IEEEtran.bst: No hyphenation pattern has been}%
\typeout{** loaded for the language `#1'. Using the pattern for}%
\typeout{** the default language instead.}%
\else
\language=\csname l@#1\endcsname
\fi
#2}}
\providecommand{\BIBdecl}{\relax}
\BIBdecl

\bibitem{9337994}
J.~Peng, W.~Shi, J.~Pang, F.~You, and S.~He, ``Iterative learning control for
  signal separation in dual-{RF} input {Doherty} transmitter,'' in \emph{Proc.
  Eur. Microw. Conf. (EuMC)}, 2021, pp. 953--956.

\bibitem{9575024}
X.~Xia, Y.~Liu, C.~Li, W.~Guo, C.~Shi, S.~Shao, L.~Lei, and Y.~Tang, ``A
  high-accuracy digital predistorter constructed by reproducing iterations of
  {ILC} with cascade architecture,'' in \emph{Proc. IEEE MTT-S Int. Microw.
  Symp. (IMS)}, 2021, pp. 446--449.

\bibitem{9322378}
A.~Fawzy, S.~Sun, T.~J. Lim, Y.~X. Guo, and P.~H. Tan, ``Iterative learning
  control for pre-distortion design in wideband direct-conversion
  transmitters,'' in \emph{Proc. IEEE Global Commun. Conf. (GLOBECOM)}, 2020,
  pp. 1--6.

\bibitem{9360062}
S.~Wang, W.~Cao, and T.~Eriksson, ``Identification methods with different
  digital predistortion models for power amplifiers with strong nonlinearity
  and memory effects,'' in \emph{Proc. IEEE MTT-S Int. Wireless Symp. (IWS)},
  2020, pp. 1--3.

\bibitem{10018540}
A.~Fawzy, S.~Sun, T.~J. Lim, Y.~Yu, C.~Yu, and Y.~Guo, ``Iterative learning
  control based digital pre-distortion for mitigating impairments in {MIMO}
  wireless transmitters,'' \emph{IEEE Trans. Veh. Technol.}, vol.~72, no.~6,
  pp. 6933--6947, June 2023.

\bibitem{7885077}
J.~Chani-Cahuana, M.~\"{O}zen, C.~Fager, and T.~Eriksson, ``Digital
  predistortion parameter identification for {RF} power amplifiers using
  real-valued output data,'' \emph{IEEE Trans. Circuits Syst. II: Exp. Briefs},
  vol.~64, no.~10, pp. 1227--1231, Oct. 2017.

\bibitem{Eun1997}
C.~Eun and E.~Powers, ``A new {Volterra} predistorter based on the indirect
  learning architecture,'' \emph{IEEE Trans. Signal Process.}, vol.~45, no.~1,
  pp. 223--227, Jan. 1997.

\bibitem{9825005}
C.~Liu, L.~Luo, J.~Wang, C.~Zhang, and C.~Pan, ``A new digital predistortion
  based on {B} spline function with compressive sampling pruning,'' in
  \emph{Proc. Int. Wireless Commun. Mobile Comput. (IWCMC)}, 2022, pp.
  1200--1205.

\bibitem{He2022}
Z.~He and F.~Tong, ``Residual {RNN} models with pruning for digital
  predistortion of {RF} power amplifiers,'' \emph{IEEE Trans. Veh. Technol.},
  vol.~71, no.~9, pp. 9735--9750, Sept. 2022.

\bibitem{7522645}
J.~Chani-Cahuana, P.~N. Landin, C.~Fager, and T.~Eriksson, ``Iterative learning
  control for {RF} power amplifier linearization,'' \emph{IEEE Trans. Microw.
  Theory Techn.}, vol.~64, no.~9, pp. 2778--2789, July 2016.

\bibitem{Pavan2016}
S.~Pavan, R.~Schreier, and G.~Temes, \emph{Understanding delta-sigma data
  converters}.\hskip 1em plus 0.5em minus 0.4em\relax Wiley \& Sons, 2016.

\bibitem{Liu2021}
L.~Liu, Y.~Ma, and N.~Yi, ``Hermite expansion model and {LMMSE} analysis for
  low-resolution quantized {MIMO} detection,'' \emph{IEEE Trans. Signal
  Process.}, vol.~69, pp. 5313--5328, Sept. 2021.

\bibitem{Morgan2006}
D.~Morgan, Z.~Ma, J.~Kim, M.~Zierdt, and J.~Pastalan, ``A generalized memory
  polynomial model for digital predistortion of {RF} power amplifiers,''
  \emph{IEEE Trans. Signal Process.}, vol.~54, no.~10, pp. 3852--3860, Oct.
  2006.

\end{thebibliography}

\end{document}